\documentclass[12pt]{article}

\usepackage[letterpaper, margin=1in]{geometry}
\usepackage[utf8]{inputenc}
\usepackage{times}
\usepackage{graphicx}
\usepackage{mathptmx}
\usepackage{amsmath}
\usepackage{dsfont}
\usepackage{amsfonts}
\usepackage{amssymb}
\usepackage[justify]{ragged2e}
\usepackage[labelfont=bf,labelsep=period]{caption}
\usepackage[style=nature,maxbibnames=20,url=false,eprint=false,isbn=false]{biblatex}
\usepackage{hyperref}
\usepackage{siunitx}
\hypersetup{
    colorlinks=true,
    linkcolor=blue,
    filecolor=magenta,      
    urlcolor=cyan,
}
\usepackage{titling}
\usepackage{soul,xcolor} 
\usepackage{authblk}
\usepackage{tabularx}

\begin{document} 

\title{Concurrent Multifractality and Anomalous Hall Response in the Nodal Line Semimetal Fe$_3$GeTe$_2$ Near Localization}

\author[1,2]{\small Subramanian Mathimalar $^*$}
\author[1]{\small Ambikesh Gupta$^*$}
\author[1]{\small Yotam Roet$^*$}
\author[3,4]{\small Stanislaw Galeski$^*$}
\author[3]{\small Rafal Wawrzynczak}
\author[5,6]{\small Mikel Garcia-Diez}
\author[3,5]{\small I\~{n}igo Robredo}
\author[3]{\small Praveen Vir}
\author[3,7]{\small Nitesh Kumar}
\author[3]{\small Walter Schnelle}
\author[8]{\small Karin von Arx}
\author[8]{\small Julia Küspert}
\author[8,9]{\small Qisi Wang}
\author[8]{\small Johan Chang}
\author[10]{\small Yasmine Sassa}
\author[1]{\small Ady Stern}
\author[11]{\small Felix von Oppen}
\author[3,5]{\small Maia G. Vergniory}
\author[3]{\small Claudia Felser}
\author[3,4]{\small Johannes Gooth}
\author[1]{\small Nurit Avraham$^\dagger$}
\author[1]{\small Haim Beidenkopf$^\dagger$}

\affil[1]{\footnotesize Department of Condensed Matter Physics, Weizmann Institute of Science, Rehovot, Israel.}
\affil[2]{\footnotesize International Institute of Information Technology, Hyderabad, 500032, India.}
\affil[3]{\footnotesize Max Planck Institute for Chemical Physics of Solids, Nöthnitzer Straße 40, 01187 Dresden, Germany.}
\affil[4]{\footnotesize Physikalisches Institut, Universität Bonn, Nussallee 12, 53115 Bonn, Germany.}
\affil[5]{\footnotesize Donostia International Physics Center, 20018 Donostia-San Sebastian, Spain.}
\affil[6]{\footnotesize Department of Physics, University of the Basque Country (UPV/EHU), 48080 Bilbao, Spain.}
\affil[7]{N. Bose National Centre for Basic Sciences, Salt Lake City, Kolkata-700106, India}
\affil[8]{\footnotesize Physik-Institut, Universität Zürich, Winterthurerstrasse 190, CH-8057, Zürich, Switzerland.} 
\affil[9]{\footnotesize Department of Physics, The Chinese University of Hong Kong, Shatin, Hong Kong, China.} 
\affil[10]{\footnotesize Department of Applied Physics, KTH Royal Institute of Technology, SE-106 91 Stockholm, Sweden.}
\affil[11]{\footnotesize Dahlem Center for Complex Quantum Systems,  Freie Universit\"{a}t Berlin, 14195 Berlin, Germany.}

\maketitle

\section*{Abstract}
\textbf{Topological states of matter exhibit unique protection against scattering by disorder. Different topological classes exhibit distinct forms and degrees of protection. Here, we investigate the response of the ferromagnetic nodal line semimetal Fe$_3$GeTe$_2$ to disorder and electronic interactions. By combining global magneto-transport with atomic-scale scanning tunneling spectroscopy we find a simultaneous onset of diverse phenomena below a common temperature scale of about 15 K: A crossover from metallic to insulating temperature dependence of the longitudinal resistivity, saturation of the anomalous Hall conductivity to its maximal value, formation of a sharp zero-bias dip in the tunneling density of state, and emergence of multi-fractal structure of the electronic wavefunction peaking at the Fermi energy. These concurrent observations reflect the emergence of a novel energy scale possibly related to the opening of a gap in the nodal line band of Fe$_3$GeTe$_2$. Our study provides overarching insight into the role of disorder, electronic interactions and Berry curvature in setting the micro- and macro-scale responses of topological semimetals.}


Many topological phases of matter host exotic electronic states at their boundaries, and these states possess a certain degree of protection against disorder \cite{zhangnatmat2017, RevModPhys.88.021004}. The mode and degree of protection varies between the different topological classes. Among these, gapped topological phases exhibit the most robust form of protection against scattering by disorder. Prime examples of bulk gapped topological classes are strong topological insulators that host an odd number of Dirac surface bands protected by time reversal symmetry \cite{RevModPhys.83.1057, Bernevig+2013}; the topological crystalline insulator that hosts an even number of Dirac surface bands protected by certain crystalline symmetries \cite{fu2011topological}; as well as the canonical quantum Hall state that hosts chiral modes at its edges \cite{cage2012quantum}. The chiral or helical nature of those boundary electronic modes protects them from backscattering and hence from localization by disorder. The topological protection is robust as long as the protecting symmetries are preserved and the disorder strength is weaker than the bulk gap \cite{beidenkopf2011spatial}. Intriguingly, in the quantum Hall state disorder localizes the bulk states and by doing so promotes protection of the chiral edge modes. 

Gapless topological phases of matter present a more restricted level of protection against disorder. Weyl semimetals are defined by bulk nodal points with Dirac-like dispersion \cite{RevModPhys.90.015001}. Correspondingly, open-contour Fermi arcs appear on their surfaces whose penetration depth  into the bulk diverges towards the surface projections of the bulk nodes. Fermi arcs thus present an interesting mixture of surface and bulk bands that affects their susceptibility to disorder and surface potentials \cite{inoue2016quasiparticle,batabyal2016visualizing,morali2019fermi,PhysRevB.97.235108,cos2_exp,cos2_theory}. In the absence of a gap, the robustness of the bulk nodal bands against disorder and scattering is quite subtle due to the finite density of states (DOS) throughout the semimetallic spectrum except at the nodal points. While the helical or chiral nature of the Dirac or Weyl bands, respectively, suppresses intra-node scattering, inter-node scattering is suppressed only by the large momentum transfer associated with such processes. Being a phase intermediate between a trivial and a topological insulator, disorder may annihilate Weyl nodes and thus transition the Weyl semimetal into either of these states \cite{PhysRevLett.115.246603}. Yet, in contrast to the two-dimensional nodal semimetal graphene, as long as node annihilation is prevented, the vanishing DOS at the three-dimensional Weyl nodes has been predicted to be robust to a finite degree of disorder \cite{PhysRevLett.107.196803, PhysRevLett.108.046602, PhysRevB.93.085103, doi:10.1146/annurev-conmatphys-033117-054037,PhysRevB.98.205134}. Such a transition from zero to non-zero DOS at a finite critical disorder strength lies beyond the Anderson localization scenario that has been otherwise considered the ultimate fate of non-interacting disordered three dimensional systems \cite{PhysRev.109.1492}. Yet, rare events were later suggested to induce nonzero DOS at lower disorder strength turning this transition into a crossover \cite{PhysRevB.89.245110, PhysRevX.6.021042, PIXLEY2021168455}.

The topological protection of nodal line semimetals against disorder is far less explored. This three-dimensional electronic phase has a nodal Dirac dispersion along two momentum directions with a continuous line of nodes along the third. Introducing a gap except at specific discrete points along the nodal line leads to the emergence of a Weyl semimetal, while fully gapping the nodal line may result in a topological insulator. The nodal line removes the distinction between inter- and intra-node scattering that pertains to Weyl semimetals. Furthermore, it was shown that the presence of the nodal line and its associated Berry curvature may support an anomalous Hall effect \cite{PhysRevB.97.161113} and weak antilocalization due to interference paths that encircle the nodal line \cite{PhysRevLett.122.196603}. A recent study predicts that also in nodal line semimetals the zero DOS would be resilient to finite disorder, concentrating the momentum-space amplitude of the electronic wave function around the nodal line and leading to multi-fractal behavior \cite{PhysRevLett.124.136405}. At a critical disorder strength, a semimetal to compressible metal transition is claimed to occur, coinciding with a multi- to single-fractality transition. In other studies this result is refuted \cite{PhysRevB.93.035138}. In type II nodal line semimetals that have a tilted Dirac spectrum, disorder is further predicted to split the nodal line into two exceptional lines with a Fermi ribbon dispersing among them \cite{PhysRevB.99.041116}. To date, these intriguing theoretical observations do not settle into a consistent framework for the response of nodal line semimetals to disorder, and were not  probed experimentally. 

\begin{figure}[ht]
\centering
\includegraphics[width=0.98\linewidth]{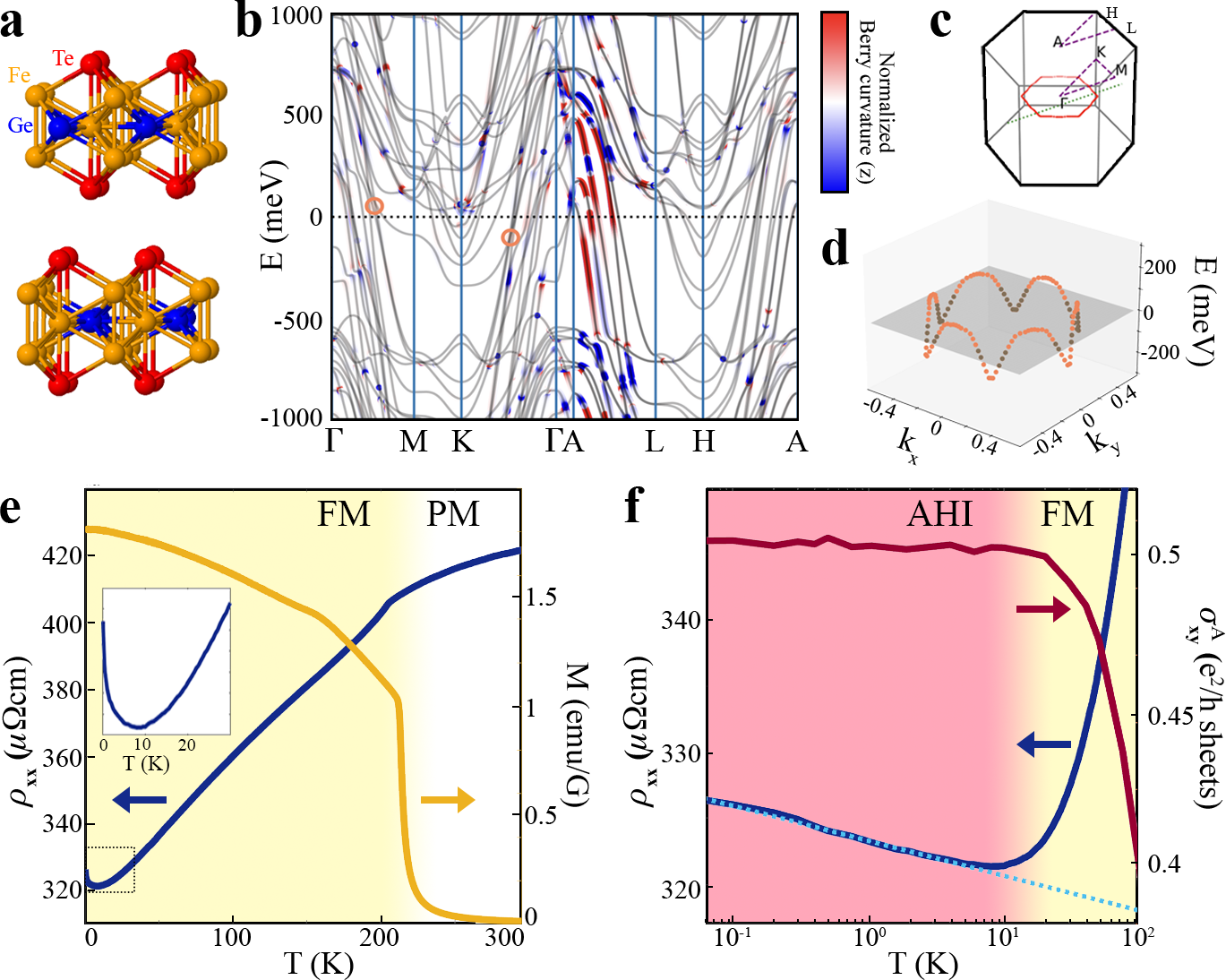}
\caption{\textbf{Magneto-electric responses of the nodal line semimetal Fe$_3$GeTe$_2$. a} Crystal structure of Fe$_3$GeTe$_2$ having three inequivalent Fe atomic sites relative to the Te cleave surface. \textbf{b} Band structure calculation from the Wannier tight binding model with the projected Berry curvature in the z direction. \textbf{c} First Brillouin zone and one highlighted nodal line (red). The purple, dashed lines are high-symmetry paths for the band structure and the dotted green line is the path for the Wilson loop calculation. \textbf{d} Energy dispersion of the nodal line in panel c and corresponding to the orange circles in panel b, which crosses the Fermi energy (gray plane). \textbf{e} The Curie temperature, $T_c=220$ K is traced in the out of plane magnetization (yellow line, right axis) as well as in the longitudinal resistivity (blue line, left axis). The inset shows a non-monotonic behavior of the longitudinal resistivity at about 15 K. \textbf{f} Below about 15 K the temperature dependence of the longitudinal resistivity (blue line, left axis) turns from metallic to insulating-like with logarithmic rise with decreasing temperature. Concurrently the anomalous Hall conductivity per layer saturates to its maximal value (red line, right axis).}
\label{fig1}
\end{figure}

Here, we investigate the response of the ferromagnetic nodal line semimetal Fe$_3$GeTe$_2$ to intrinsic disorder in the form of Fe vacancies and to electronic interactions. Fe$_3$GeTe$_2$ is a Van der Waals compound, whose crystal structure is shown in Fig.\ref{fig1}a. It has a high Curie temperature of about T$_c$=220 K. Room temperature ferromagnetism was established for thin-film samples under ionic gating \cite{doi:10.1021/acs.nanolett.8b02806}. Previous ab initio calculations classified Fe$_3$GeTe$_2$ as a nodal line semimetal whose nodal line is slightly gapped by spin-orbit interaction and weakly disperses along the out of plane K-H direction crossing the Fermi energy \cite{kim2018}. In addition to the nodal line band it hosts several other bands that cross the Fermi energy rendering it a metal. This nodal line has been shown to be an effective source of a large anomalous Hall response \cite{kim2018}. Furthermore, as the band crossing is present even in a bilayer, mechanically exfoliated ultrathin crystals may realize the quantized anomalous Hall effect. We investigate this intriguing compound by a combination of global magneto-transport and atomic scale scanning tunneling microscopy (STM) and spectroscopy mappings. Our complementary macro- and micro-scale approaches suggest a relation between the strong anomalous Hall response accompanied by an upturn in longitudinal resistivity and a zero-bias suppressed tunneling DOS (TDOS) accompanied by multi-fractal structure of the electronic wavefunction, all forming below a common temperature scale of about 15 K in bulk Fe$_3$GeTe$_2$.

We first revisit the ab initio calculation of the band structure of Fe$_3$GeTe$_2$, shown in Fig.\ref{fig1}b along high symmetry cuts of the Brillouin zone (Fig.\ref{fig1}c). We used the Vienna Ab Initio Simulation Package (VASP) \cite{vasp} implementation of DFT for the computations (see section \ref{sm-dft} for details). Remarkably, we find numerous nodal lines in the in-plane direction in addition to the out-of-plane dispersing one that has been previously reported \cite{kim2018}. Even though spin degeneracy is lifted by the ferromagnetic order, the mirror symmetry of the $k_z=0, \pi$ planes prevents gapping of these nodal lines. They remain gapless also in the presence of spin orbit coupling. At least three of the those nodal lines, as the one shown in Fig.\ref{fig1}d, disperse across and in the vicinity of the Fermi energy that we identify by comparison of the band structure found in ab initio calculation to that imaged in angular resolved photo-emission spectroscopy (Fig.\ref{figs-arpes}). Our calculations further find that all the nodal bands are type II in nature. Weakly gapped nodal bands contribute to the total Berry curvature,  rendering Fe$_3$GeTe$_2$ a favorable compound for detection of a quantum anomalous Hall response. 

We next describe our main magneto-transport findings. Magnetization measurements clearly resolve the ferromagnetic transition at 220 K (Fig.\ref{fig1}e, yellow line). This value of the Curie temperature in Fe$_{3-x}$GeTe$_2$ suggests \cite{kim2018} an Fe vacancy concentration of  $x\approx$0.12 in our samples. Onset of magnetization is accompanied by hysteresis (Fig.\ref{figs-mh}). The slanted magnetization loops have been attributed to the presence of ferromagnetic domains \cite{doi:10.1021/acs.nanolett.8b02806, doi:10.1063/1.4961592}. The longitudinal resistivity, $\rho_{xx}$, shows metallic behavior with a phonon-dominated temperature dependence (Fig.\ref{fig1}e, blue line) across a wide temperature range. In this study we particularly focus on a peculiar rise in the longitudinal resistivity that occurs at low temperatures, highlighted in the inset of Fig.\ref{fig1}e. Below about 15 K the metallic temperature dependence turns into an insulator-like logarithmic increase (fit included as dotted line in Fig.\ref{fig1}f). The logarithmic rise may be attributed to a Kondo transition. Evidence of a Kondo transition has been reported before in Fe$_3$GeTe$_2$ though at higher temperature \cite{zhang2018, Zhao2021, Bao2022}. Alternatively, it may signify the combined effects of  disorder and electronic interactions \cite{PhysRevLett.44.1288} as further elaborated below. 

The magnetic field dependence of the transverse Hall resistivity, $\rho_{xy}$, shows a pronounced anomalous Hall contribution that offsets the linear magnetic field dependence of the weak-field Hall signal (see Fig.\ref{figs-rsh}a). From the longitudinal and Hall resistivities, we estimate the strength of disorder scattering by extracting $k_Fl$, where $k_F$ is the Fermi wavenumber and $l$ is the electronic mean free path (See section \ref{mag-trans}). We find a value of $k_Fl\approx$5-10. However, there are multiple electronic pockets that cross the Fermi energy as suggested by the ab initio calculation (see Fig.\ref{fig1}b). If all bands had the same Fermi wavelength and contributed equally to the conductivity, this would suggest that they all have $k_Fl\approx 1$. In view of the varying Fermi wavelengths, we expect values on the order of 1 and bounded by 5 per electronic band, indicating a disordered metallic state as already indicated by the large residual resistivity.  

We extract the anomalous Hall conductivity from the intercept of the linear extrapolation of $\sigma_{xy}$ to zero magnetic field (see Fig.\ref{figs-rsh}a). The temperature dependence of the anomalous Hall conductivity, $\sigma_{xy}^A$, is shown in Fig.\ref{fig1}f (red line). Intriguingly, concurrent with the onset of insulator-like behavior in the longitudinal resistivity we find saturation of the anomalous Hall conductivity. We emphasize that this has no counterpart in the magnetization and thus signifies saturation of the intrinsic anomalous Hall conductivity (see analysis in Fig.\ref{figs-sat}). While the value we found per Van der Waals layer (of thickness $a_z$=0.8 nm) is not quantized in units of quantum of conductance, $e^2/h$, it is of order unity in the range of 0.5-1.1$\pm0.1$ $a_z e^2/h$ in the three samples we have characterized (Fig.\ref{figs-rst}c). Saturation around 15 K was consistently observed across different samples despite their distinct longitudinal resistivities at high temperatures (Fig.\ref{figs-rst}).

\begin{figure}[ht]
\centering
\includegraphics[width=0.5\linewidth]{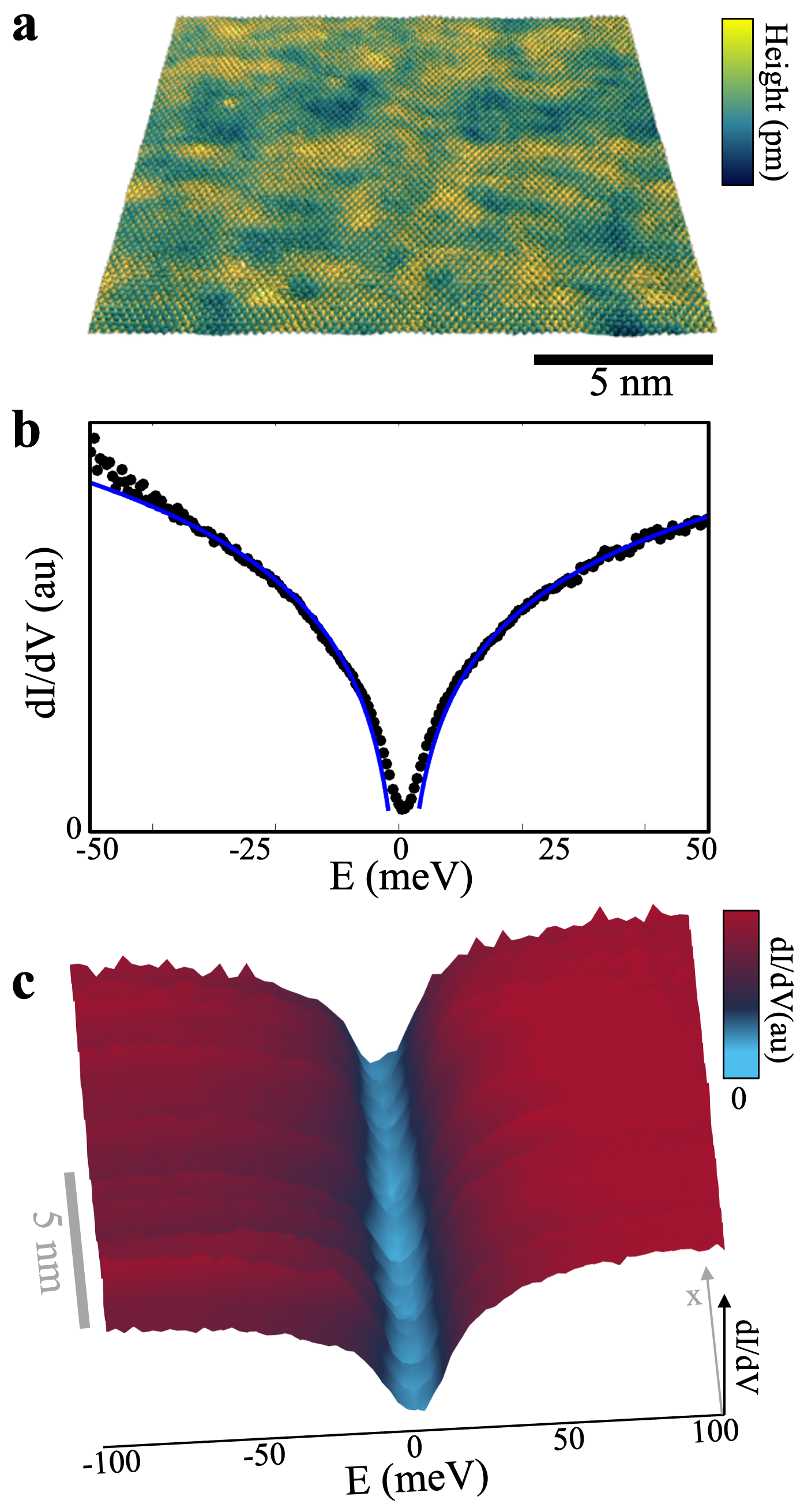}
\caption{\textbf{Zero bias dip in the tunneling density of states. a} Representative topographic image of the Te terminated surface showing disorder pattern attributed to subsurface Fe vacancies. \textbf{b} Typical dI/dV spectrum taken at 400 mK with ac excitation of $V_{ac}=$1 meV showing a pronounced zero bias dip. The blue line is a fit to Altshuler-Aronov model.
\textbf{c} Spectroscopic line cut measured across the sample surface showing robust spectral features with mild spatial fluctuations within a single field of view.}
\label{fig2}
\end{figure}

We now turn to spectroscopic characterization by STM of the electronic states at low temperatures. In topography, shown in Fig.\ref{fig2}a, we find clear indication of abundant Fe vacancies, in agreement with previous STM reports \cite{zhang2018}. Fe$_3$GeTe$_2$ commonly hosts Fe vacancies as a main source of native disorder. Since there are three different Fe sites relative to the cleaved Te surface in the top-most Van der Waals layer and we image the terminating Te layer (see structure in Fig.\ref{fig1}a), the vacancies do not appear as single-atomic but rather spread over several atomic sites. They thus appear as a disorder landscape with typical scale of about 1 nm.

Surprisingly, the differential conductance (dI/dV)  displayed in Fig.\ref{fig2}b, observed across the sample's surface, bears no resemblance to the metallic behavior anticipated based on the single-particle ab initio calculations shown in Fig.\ref{fig1}b. Instead we measure a sharp and deep suppression of the local tunneling DOS (TDOS) whose minimum is pinned to the Fermi energy as signified by zero bias in STM measurements (see also Fig.\ref{figs-soft}a). The existence of a zero bias suppression in the spectrum is robust across the surface apart from variations in the residual zero-bias conductance. The line cut displayed in Fig.\ref{fig2}c  is representative of regions with almost complete suppression of the TDOS at the Fermi energy. The curvature of the energy dependent dI/dV spectrum about zero bias may suggest slight gapping (see Fig.\ref{figs-soft}c). In other regions, we have also detected less pronounced zero-bias suppressions of the local TDOS (see, e.g., Fig.\ref{figs-soft}b), but the energy dependence is similar. Importantly, some level of suppression is consistently observed throughout the sample. 

\begin{figure}[ht]
\centering
\includegraphics[width=0.75\linewidth]{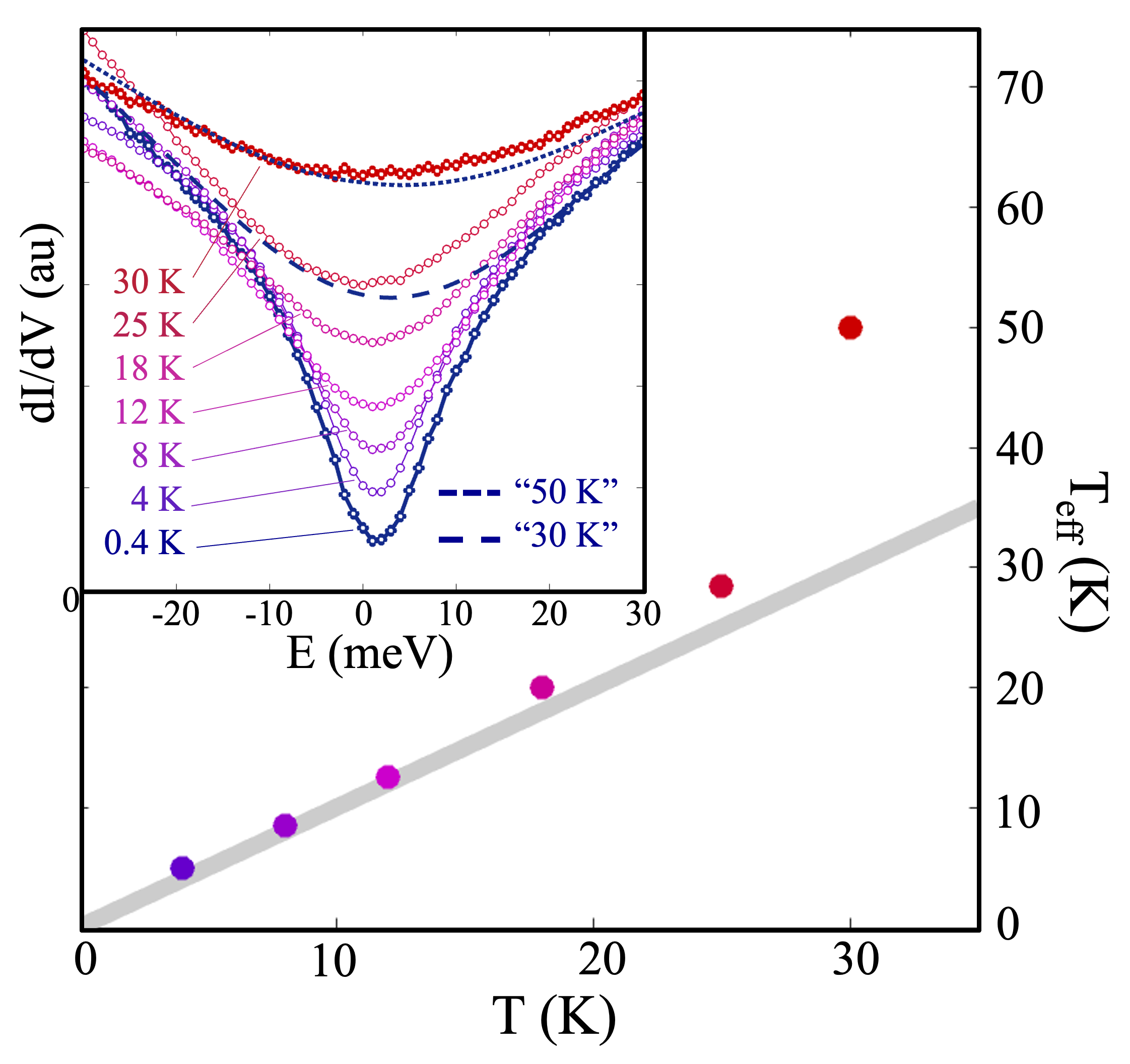}
\caption{\textbf{Temperature evolution of the zero bias dip}. The inset shows a series of dI/dV spectra measured at rising temperature from 400 mK to 30 K all taken with DC bias of $V_B=$50 meV and AC excitation of $V_{ac}=$1 meV. The dashed line shows the result of thermally broadening the 0.4 K dI/dV by 30 K, and the dotted line the result of broadening with an effective temperature of $T_{eff}=$50 K, that fits well the 30 K measured dI/dV. The main panel shows the fitted effective temperatures for dI/dV spectra taken between 4-30 K (see fits in Fig.\ref{figs-teff}). The solid line has a slope of $T_{eff} \propto T$ passing through the lowest data point.}
\label{fig3}
\end{figure}

Figure \ref{fig3} examines the temperature evolution of this unexpected suppression. The inset shows dI/dV spectra averaged over the same field of view as temperature is varied  between 400 mK and 30 K. The dip in dI/dV gradually diminishes as temperature is raised. Thermal broadening entering through the Fermi-Dirac distribution would smoothen the zero-bias suppression of the TDOS found at low temperatures. We can account for this effect by convolving the measured lowest temperature spectrum with the derivative of the Fermi-Dirac distribution  \cite{Devoret1992}. Yet, when doing so, e.g., for the 30K data, we find that the resulting curve, given by the dashed line in the inset of Fig.\ref{fig3}, is far from capturing the full effect of the elevated temperature (red curve). An unrealistic effective temperature of $T_{eff}=$50 K (dotted line) would be needed to obtain a reasonable fit to the measured $T=$30 K spectrum. The main panel of Fig.\ref{fig3} shows the effective temperatures extracted from fitting all elevated temperature dI/dV spectra by thermally broadening the lowest temperature spectrum measured at 0.4 K (see individual fits in Fig.\ref{figs-teff}). As temperature rises, the effective temperature progressively deviates from the $T_{eff} \propto T$ trend shown by a solid line. This deviation signifies that the zero-bias suppression of the TDOS diminishes faster with temperature than predicted by thermal broadening.

Such a dip in the low-energy TDOS is characteristic of the Altshuler-Aronov response of a metal to disorder and interactions \cite{AltshulerAronov1979}. Because of the Van der Waals nature of the compound ($\rho_c/\rho_{ab} \approx$ 15 \cite{wang2017}), we approximate the electronic dynamics as two dimensional, while retaining the Coulomb interaction as three dimensional. Appropriately adapting the conventional calculation \cite{AltshulerAronov1979}, we find a logarithmic correction to the TDOS (see section \ref{sm-aa} for details), 
\begin{equation} 
    \frac{\nu(\varepsilon)}{\nu_0} = 1-\frac{\kappa d}{k_Fl} \log{\frac{\hslash D\kappa^2}{\varepsilon}}, \;\;\;\; \kappa{}^2 = 4 \pi e^2 \frac{\nu_{2D}}{d}, 
\end{equation}
where $\nu_0$ is the unperturbed DOS, $\kappa$ is the two dimensional Thomas-Fermi screening length, $D$ is the diffusion constant, and $d$=2$a_z$=1.6 nm is the unit cell thickness. This result is obtained perturbatively in both disorder ($k_Fl\gg 1$) and interaction. As demonstrated in Fig.\ref{fig2}b by the solid blue line, the energy dependence captures well the dI/dV spectra we find in spectroscopy. From the fit, we extract $\kappa d/k_Fl=0.2$. Using $k_Fl\approx$5-10  this yields $\lambda_{TF}=2\pi/\kappa\approx$5-10 nm while for metals the typical value is less than a nanometer, consistent with the low density of electrons in Fe$_3$GeTe$_2$ \cite{doi:10.1021/acsnano.2c00512}. From $\hslash D\kappa^2$=274 meV as obtained from the fit, we find  $D\approx$ 3-10 cm$^2$/s. This is about an order of magnitude smaller than in typical metals, as might be expected from the high levels of disorder in our sample.    

An alternative origin of the TDOS suppression is the Kondo effect, that we discuss in section \ref{sm-kondo} of the supplementary material  (see comparison in Table \ref{tbl-kondo-aa}). A logarithmic increase in longitudinal resistivity is indeed characteristic of a Kondo transition. The zero bias dip in dI/dV can be well fitted with a Fano line shape, though in the extreme limit of tunneling predominantly to the screening metal. The temperature dependence yields a crossover with a Kondo temperature of $T_K\approx$ 30 K. We focus here on the disorder origin because a strong ferromagnet is not a natural state to host magnetically fluctuating localized spins. 

\begin{figure}[ht]
\centering
\includegraphics[width=1\linewidth]{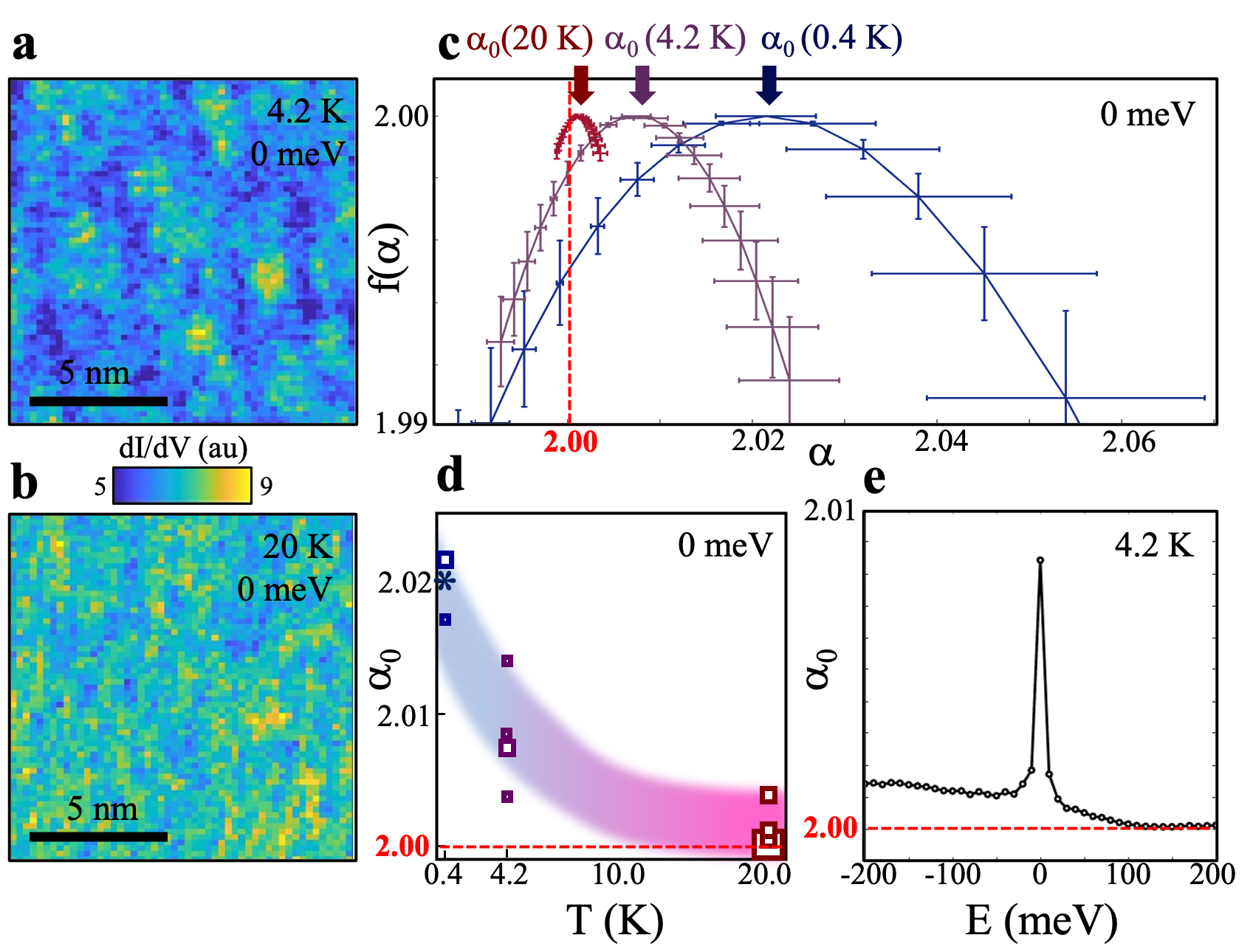}
\caption{\textbf{Multi-fractal structure of the wavefunction. a(b)} Spectroscopic mapping of the spatial fluctuations in zero bias dI/dV at 4.2 K (20 K) \textbf{c} Multi-fractal analysis of dI/dV maps showing onset of multi-fractality as the sample is cooled from 20 K to 400 mK (parabolic maxima shift away from the physical dimension of 2.00 and broaden). All maps were taken at a DC bias of $V_b=$200 meV and AC excitation of $V_{ac}=$2 meV. \textbf{d} Extracted leading fractal dimension, $\alpha_0$, of ten spectroscopic dI/dV maps showing clear trend (symbol size represents the map resolution - 32X32, 64X64, 128X128). The value extracted from the longitudinal resistivity is marked by asterisk. \textbf{e} the leading fractal dimension sharply peaks at the Fermi energy signifying it results from electronic interactions.}
\label{fig4}
\end{figure}

To corroborate the importance of disorder, we analyze the spatial structure of the electronic wavefunctions. Indeed, the local TDOS in Fe$_3$GeTe$_2$ exhibits strong spatial fluctuations in the presence of the Fe vacancies, as mapped at zero bias in Fig.\ref{fig4}a and b for temperatures of 4.2 K and 20 K, respectively. We note, that throughout our spectroscopic mappings we did not identify an extended structure that could suggest a spectroscopic response to a ferromagnetic domain wall so we conclude the fluctuating TDOS is dominated by the response to point disorder. We characterize these fluctuations by multi-fractal analysis. They are captured by an infinite set of critical exponents describing the scaling of the moments of the wave-function amplitude. multi-fractality occurs in several Anderson localization models, and is usually associated with critical points of the model. Accordingly it is predicted to occur in three-dimensional models of non-interacting electrons which exhibit a metal to insulator localization transition \cite{janssen1994multifractal,RevModPhys.67.357,Evers2008}. In two dimensions, it is absent for symmetry classes which localize at any disorder strength \cite{PhysRevLett.42.673}, but present for integer quantum Hall states, hosting multi-fractal wave functions at the plateau transitions \cite{RevModPhys.67.357,PhysRevLett.90.056804}, as well as in quantum spin Hall systems \cite{PhysRevLett.98.256801}. Experimentally, multi-fractal wave function structures were previously demonstrated by STM mappings in a few electronic systems, including quantum Hall states \cite{PhysRevLett.90.056804}, Mn-doped GaAs which undergoes a metal-insulator transition with Mn doping \cite{Richardella2010}, disordered two-dimensional films \cite{PhysRevResearch.3.013022}, and superconducting systems \cite{zhao2019,doi:10.1021/acs.nanolett.0c01288}. In general STM probes the surface DOS, whose scaling and criticality may differ from that of three-dimensional bulk states \cite{PhysRevLett.96.126802}.

From the measured dI/dV maps we calculate the singularity spectrum, $f(\alpha)$. It gives the fractal dimension of the set of points, $r$, for which the local DOS scales as $L^{-\alpha}$ with $L$ being the system size \cite{PhysRevA.33.1141}. In a discretized system, as provided by STM measurements, the number of such pixels scales as $L^{f(\alpha)}$. In a nonfractal or monofractal systems, the singularity spectrum would consist of a narrow peak at the physical or fractal dimension, respectively. A multi-fractal system is characterized by a broad distribution ranging over a continuum of values and peaking away from the physical dimension. 

The extracted singularity spectra for Fe$_3$GeTe$_2$ at various temperatures are shown in Fig.\ref{fig4}c. At 20 K the dI/dV map is nonfractal as $f(\alpha)$ peaks sharply close to the physical dimension at $\alpha\approx 2$. However, at the lower temperatures of 4.2 K and 400 mK, multi-fractality progressively onsets as the peak broadens and its maximum, $\alpha_0$, shifts away from the physical dimension of two. The maximal deviation that we find is $\alpha_0 \approx$2.02. For non-interacting quasi two-dimensional systems with $\sigma_{xy}\ll\sigma_{xx}$, the expected deviation from the physical dimension can be related to the longitudinal conductance through $\alpha_0-2=\frac{2 e^2}{\pi\beta_0 h\sigma_{xx}}$ \cite{efetov_1996}, where $\beta_0 =2$ for a unitary system with broken time reversal symmetry. For $\rho_{xx}=$ 325 $\mu\Omega cm$ as measured for our system, we obtain $\alpha_0 \approx$2.02 (marked by an asterisk in Fig.\ref{fig4}d), consistent with what we extract from the dI/dV maps (see supplementary \ref{supp-frac-r} for more details). We note that in previous studies of systems that undergo a full metal to insulator transition the multi-fractal spectrum peaked at about 2.1 \cite{Richardella2010} in the most extreme cases and at less than 2.01 \cite{doi:10.1021/acs.nanolett.0c01288} elsewhere. In Fig.\ref{fig4}d, we present a collection of ten distinct  maps we have analyzed showing the onset of multi-fractality with lowering temperature below 15 K.

Multi-fractality is most pronounced at the Fermi energy, suggesting that it is driven by electronic interactions. For this we apply multi-fractal analysis to a dI/dV map taken at 4.2 K over a broad range of energies. In Fig.\ref{fig4}e we follow the deviation of the maximum, $\alpha_0$, of the singularity spectrum from the physical dimension of two, as a function of bias voltage. We find that this deviation peaks sharply at the Fermi energy and shifts swiftly back towards the physical dimension at energies away from the Fermi energy. The fate of disorder-induced multi-fractality in the presence of interactions is not fully explored theoretically at present \cite{amini2014, PhysRevLett.111.066601, PhysRevB.97.155105}. We also note that the lifetime of electrons and holes injected in STM experiments at finite bias shortens with increasing energy. Relaxation may thus also contribute to the energy dependence as dephasing was found to suppress multi-fractality \cite{BURMISTROV20111457}.  

In summary, we report here four observations that onset below a common temperature scale of about 15 K: A longitudinal resistivity that goes from decreasing to logarithmically increasing with decreasing temperature, saturation of the anomalous Hall conductivity to its maximal value, suppression of the dI/dV spectrum at the Fermi energy, and observed multi-fractality of the structure of the electronic wavefunction. The measured values of the resistivity that indicate values of $k_Fl$ close to one, suggest that disorder plays an important role. The observed anomalous Hall effect and its temperature dependence suggest the presence of Berry curvature close to the Fermi energy in accordance with these bands being nodal-line semi-metallic bands. Even though STM is a surface-sensitive probe while transport probes the bulk, we contrast the local and global observations both because of the van der Waals nature of the Fe$_3$GeTe$_2$ and because of its high impurity content that breaks translational symmetry throughout the bulk.

The common crossover temperature suggests an underlying energy scale of about 1 meV. We postulate gapping of the nodal lines that cross the Fermi energy, possibly by disorder breaking the mirror symmetry that protects them. The gapping of the nodal line introduces Berry curvature of opposite signs above and below the gap. Wherever the chemical potential is within the gap, the anomalous Hall conductivity will be determined by the Berry curvature below the gap at low temperature, and will average to zero at temperatures larger than the gap. The gapping of the nodal-line band suppresses the DOS and increases the resistance, but this effect is mild due to the presence of other bands. Disorder and interactions further lead to a suppression of the TDOS at the Fermi energy due to the Altshuler-Aronov mechanism, which will also contribute to the increase of the resistivity with decreasing temperature. The disorder is not strong enough to lead to localization-dominated transport, but is manifested in the multi-fractal structure of the electronic wavefunction. Even though the band structure description is not fully supported in the presence of strong disorder, our findings show that the Berry curvature it imposes is still effective in inducing a strong anomalous Hall response. Similar robustness has been recently observed in amorphous \cite{corbae2023observation} and in non-crystalline \cite{khan2025surface} topological matter. Our findings call for continued investigation of the interplay between multi-fractality and Berry curvature as nodal line semimetallic bands undergo localization in response to disorder and interactions.

\printbibliography
\clearpage

\clearpage

\clearpage

\clearpage

\clearpage
\section*{Methods}
\subsection*{STM measurements}
The measurements were performed in a commercial Unisoku STM. The Pt-Ir tips were characterized in a freshly prepared Cu(111) single crystal. This process ensured a robust tip with reproducible results across different cleaves and samples from different batches. All the $dI/dV$ measurements were taken using standard lock-in techniques.

\subsection*{ARPES measurements}
The ARPES experiments were carried out at the SIS beamline at the Swiss Light source. The Fe$_3$GeTe$_2$
crystals were cleaving in-situ using a standard top-post. Data were recorded at a temperature of 15~K and using 50~eV circular polarized photons.

\subsection*{Single crystal preparation}
Single crystals of Fe$_3$GeTe$_2$ were grown with the chemical vapor transport (CVT) method. Fe powder, Ge powder and Te powder were taken into a stoichiometric ratio (3:1:2). Ge and Te powder were obtained by grinding the bulk pieces. The powders were mixed homogeneously and vacuum sealed in a 10 cm long quartz tube along with a small amount of I2 (~ 5 mg/cm$^3$). A two-zone furnace with 750 $^o$C as the hot zone and 650 $^o$C as the cold zone was used for the growth. The reaction was kept at these temperatures for 4 weeks; after that, the furnace was switched off, which allowed a normal cooling to room temperature. The crystals were found to be grown at the cold zone and hexagonal plate-shaped.

\subsection*{Magnetotransport measurements}
To avoid contact resistance, only four-terminal measurements were carried out. The longitudinal $\rho_{xx}$ and Hall resistivity $\rho_{xy}$ were measured in a Hall-bar geometry. For the main measurements we have used the Keithley 6221 current source with the Keithley 2182A nanovoltmeter. For increased reliability the DC measurements were performed in the 'Delta mode'. In addition to confirm reliability of the results some measuremnts have been repeated using a standard Lock-in technique (Zurich instruments MFLI ) for comparison. In order to avoid effects of self-heating at low temperatures we have performed I-V characteristics in order to make sure all measurements were done in the linear regime. The applied current never exceeded 10 $\mu A$.

\section*{Data availability}
The data that support the plots within this paper and other findings of this study are available from the corresponding authors upon reasonable request.

\section*{Acknowledgement}

Research at Weizmann and Berlin was supported by Deutsche Forschungsgemeinschaft through CRC 183 (projects A02, A04, and C03). M.G.V., M.G.D. and I.R. acknowledge support to the Spanish Ministerio de Ciencia e Innovacion (grant PID2022-142008NB-I00), partial support from European Research Council (ERC) grant agreement no. 101020833 and the European Union NextGenerationEU/PRTR-C17.I1 and the IKUR Strategy under the collaboration agreement between Ikerbasque Foundation and DIPC on behalf of the Department of Education of the Basque Government. C.F. and M.G.V. acknowledge funding from the German Research Foundation (DFG) and the Austrian Science Fund (FOR 5249 - QUAST). I. R. acknowledges funding supported by the Ministry of Economic Affairs and Digital Transformation of the Spanish Government through the QUANTUM ENIA project call – Quantum Spain project, and by the European Union through the Recovery, Transformation and Resilience Plan – NextGenerationEU within the framework of the Digital Spain 2026 Agenda. KvA, JK, QW and JC acknowledge support from the Swiss National Science Foundation. 
We acknowledge the Paul Scherrer Institut, Villigen, Switzerland for provision of synchrotron radiation beamtime at the SLS beamline. Y.S. acknowledges funding from the Wallenberg Foundation through the grant 2021.0150.

\section*{Author contribution}
SM, AG, YR conducted the STM experiment and data analysis, SG, RW conducted the magneto-transport experiment and data analysis, MGD, IR, MGV did the band structure calculations, PV, NK, WS grew the material, KvA, JK, QW, JC, YS conducted the ARPES experiment, AS, FvO contributed the theoretical modeling, CF, JG, NA, HB conceived the study, SM, AG, , SG, HB, NA wrote the paper with contributions from all authors.

\section*{Competing interests}
The authors declare no competing interests.

\section*{Supplementary information}

\renewcommand{\figurename}{\textbf{Fig.}}
\renewcommand{\thefigure}{S\arabic{figure}}
\setcounter{figure}{0}  
\renewcommand{\thesection}{S\arabic{section}}

\title{Supplementary Information for \\ Concurrent Multifractality and Anomalous Hall Response in the Nodal Line Semimetal Fe3GeTe2 Near Localization}

\maketitle

\begin{refsection} 
\section{DFT calculation} \label{sm-dft}
In the absence of magnetism, the parent space group of the crystal is $P6_3/mmc (\#194)$. Since inversion and time-reversal symmetry (TRS) are both present, all bands are doubly degenerate (spin degenerate), even in the presence of spin-orbit coupling. Therefore, the irreps of the double space group are at least two-dimensional and always of even dimension at all points in the Brillouin zone (BZ).

With the onset of FM order, the symmetry is lowered to the magnetic space group P$6_3$/mm'c' (\#194.270). In particular, all the twofold axes lying on the xy-plane and their associated mirror planes now carry an additional time-reversal, TRS not being present on its own. Consequently, the spin degeneracy is generally lifted, with the exception of symmetry enforced degeneracies at high-symmetry points, lines or planes. Due to this, the nodal lines that are still present in the material must be confined to the only two planes which are invariant under the mirror plane $m_z$, located at $k_z = 0, \pi$. The states on these two planes have well-defined eigenvalues $\pm i$ under this symmetry operation, which protects crossings between bands with opposite eigenvalue, giving rise to symmetry-protected nodal lines.

We performed spin-polarized density functional theory (DFT) calculations \cite{PhysRevB.62.11556} using the GGA \cite{gga-pbe} exchange-correlation approximation and the DFT-D3 \cite{dft-d3} with Becke-Johnson damping function implemented in VASP \cite{vasp} to include Van der Waals interaction between layers in the bulk of the material. The plane-wave energy cutoff was set to 600 eV and the ground state was computed over a Gamma-centered $11\times 11 \times 7$ grid as required by the unit cell dimensions. The converged magnetic moments in the $z$ direction for the Fe atoms (2.3 and 1.5 Bohr magnetons depending on the Wyckoff position) agree with the experimental evidence \cite{magnetic_moments}. The Irrep Python package \cite{irrep} was used to extract the symmetry properties of the computed band structure. Based on these calculations, we constructed maximally localized Wannier functions as implemented in Wannier90 \cite{wannier90} and derived an interpolated effective tight-binding (TB) Hamiltonian in the Wannier function basis from initial projections onto $d$ orbitals for the Fe atoms and $p$ for all Te and Ge. Wilson loop and band gap calculations are performed using this TB Hamiltonian with routines implemented in WannierTools \cite{wanniertools} and the Berry curvature calculations were done with the WannierBerri \cite{wannierberri} post-processing package.

\begin{figure}
    \centering
    \includegraphics[width=0.98\linewidth]{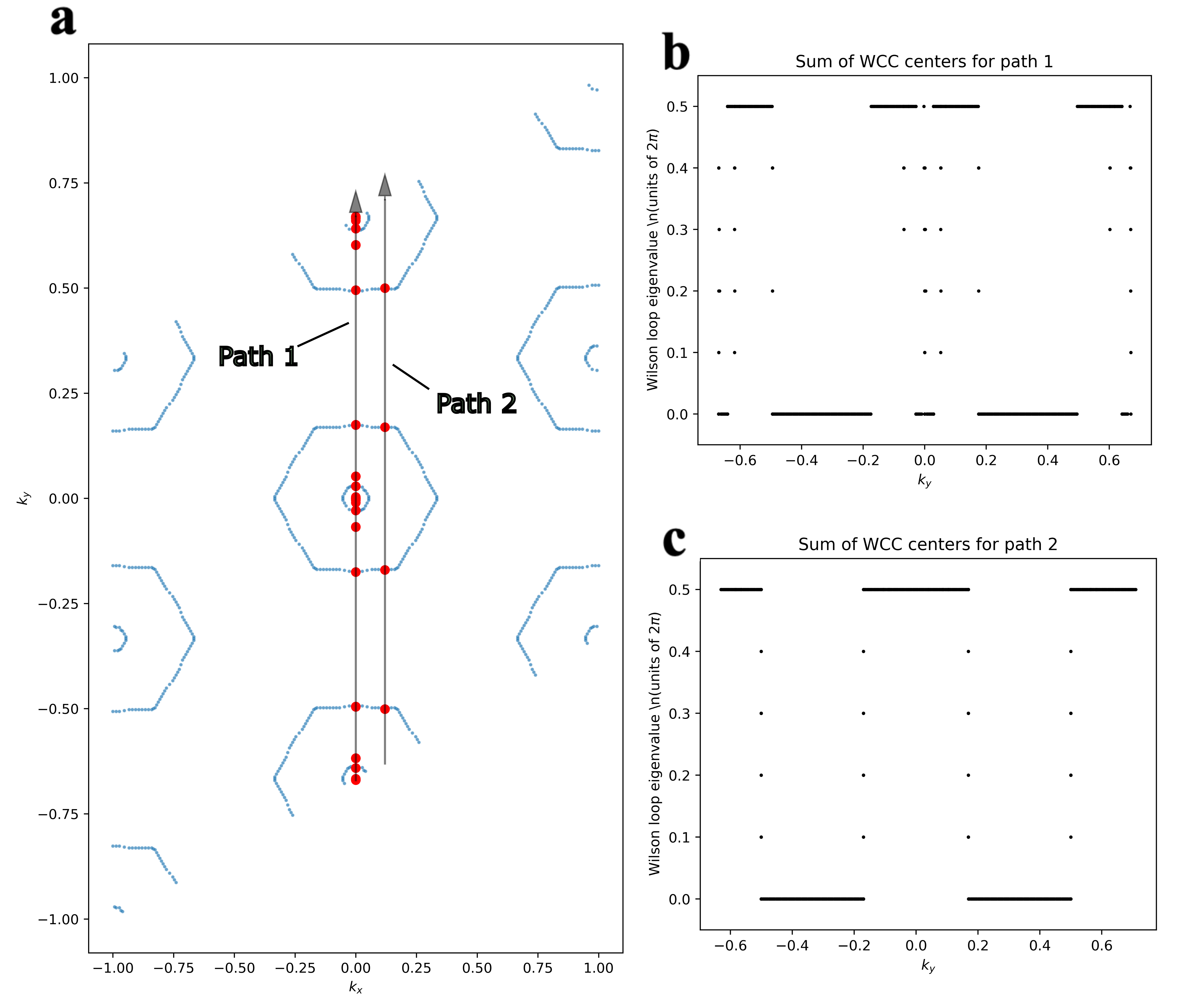}
    \caption{\textbf{Wilson loop calculations that identify the nodal line in Fig.\ref{fig1}.} \textbf{a} Nodal line at the $k_z=0$ plane between the two selected bands in Fig.~\ref{fig1}. The arrows represent the path over which the Wilson loops are calculated, integrating over $k_z$ and the dots that intersect the gapless points mark the place where the Wilson loop eigenvalue jumps by $\pi$, identifying the structures shown as nodal lines. \textbf{b}  Wilson loop eigenvalue of the lower band (modulo $2\pi$) over path 1 in (a) which identifies a nodal line for the two selected bands, not shown for the sake of clarity. \textbf{c} Wilson loop eigenvalues of the lower band for path 2 in (a), corresponding to the nodal line shown in Fig.\ref{fig1}. }
    \label{fig:supp-wilson}
\end{figure}

\begin{figure}[ht]
\centering
\includegraphics[width=0.98\linewidth]{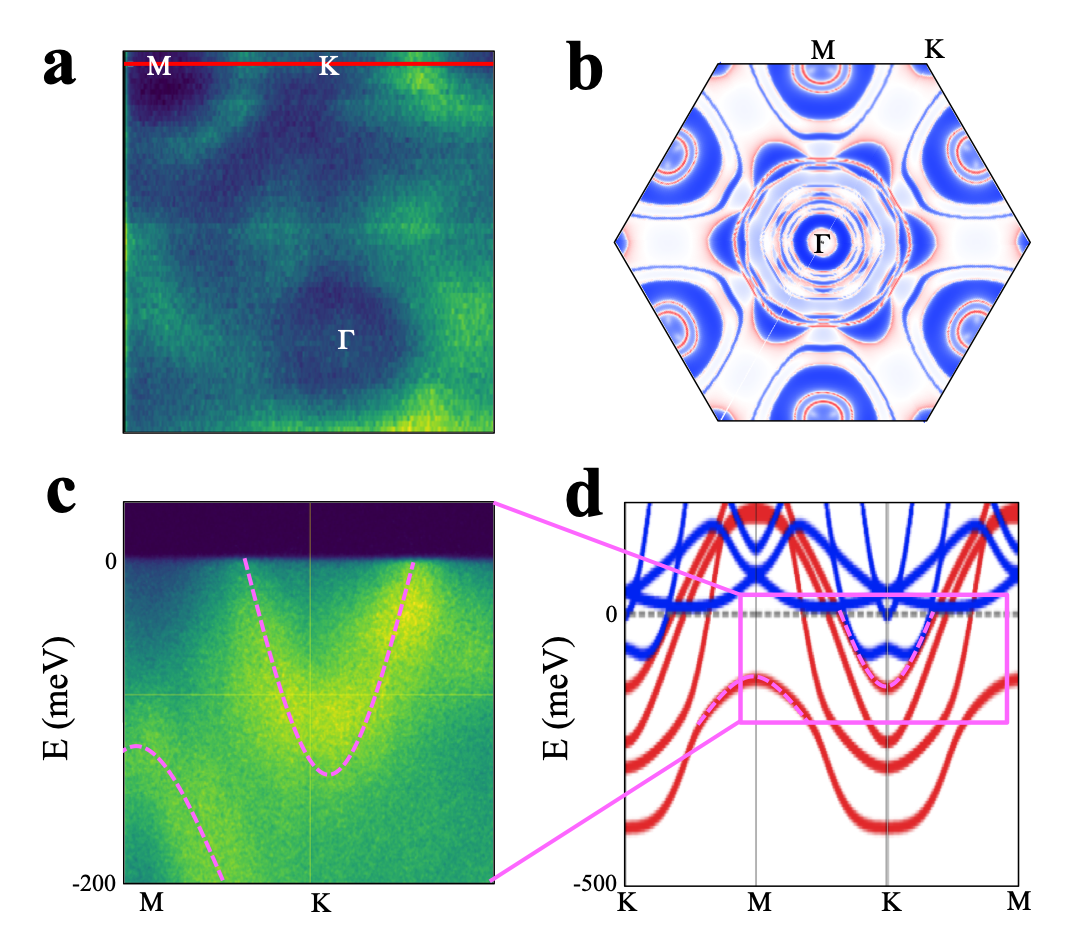}
\caption{\textbf{Comparison between ARPES and DFT. a} ARPES Fermi surface measured at 15.5 K using photon energy of 50 eV. \textbf{b} Corresponding DFT calculation of the Fermi surface showing similar pocket structure. \textbf{c} Energy-momentum cut of the ARPES data along the M-K line (red line in a). \textbf{d} Corresponding DFT calculation of the band structure along high symmetry directions.}
\label{figs-arpes}
\end{figure}

We show the resulting calculations from the Wannier TB model in Fig.\ref{fig1}. In particular, in Fig.\ref{fig1}b we find the Berry curvature in the z direction projected onto the band structure (the other two components are zero due to the $m_z$ mirror symmetry). We performed an exhaustive search of gapless points in the band structure over a window around the Fermi level to investigate the presence of nodal lines. Surprisingly, we found a rather large number of them, among which we chose to highlight one example crossing the Fermi energy. Figure \ref{fig1}c depicts the first Brillouin zone (BZ) with the high-symmetry points located on the planes where the crossing of energy levels is protected by the $m_z$ mirror symmetry (purple, dashed lines). The red line shows the nodal line in reciprocal space, which lies on the plane $k_z = 0$. Figure \ref{fig1}d shows the dispersion of this nodal line, which crosses the Fermi energy, represented by the gray plane. The points where this line intersects the computed band structure are marked on Fig.\ref{fig1}b with orange circles. Finally, to unequivocally detect whether some crossing constitutes a nodal line or not, we computed the Wilson loop eigenvalues. These calculations are done by choosing a line on the $k_z = 0$ plane and integrating over the $k_z$ direction. The Wilson
loop eigenvalues undergo a $\pi$ jump when a nodal line is crossed as it is quantized in units of $\pi$ due to the $m_z$ mirror symmetry \cite{cos2_theory}, as shown in Fig.\ref{fig:supp-wilson}.

\section{dI/dV suppression}
The extent of suppression in dI/dV around zero bias varies across the sample surface. In some regions it is complete as shown in Fig.\ref{fig2}. In other it appears as a partial suppression. We could not find any correlation between the local distribution of Fe vacancies as seen in topography and the extent of dI/dV suppression. 

\begin{figure}[ht]
\centering
\includegraphics[width=0.98\linewidth]{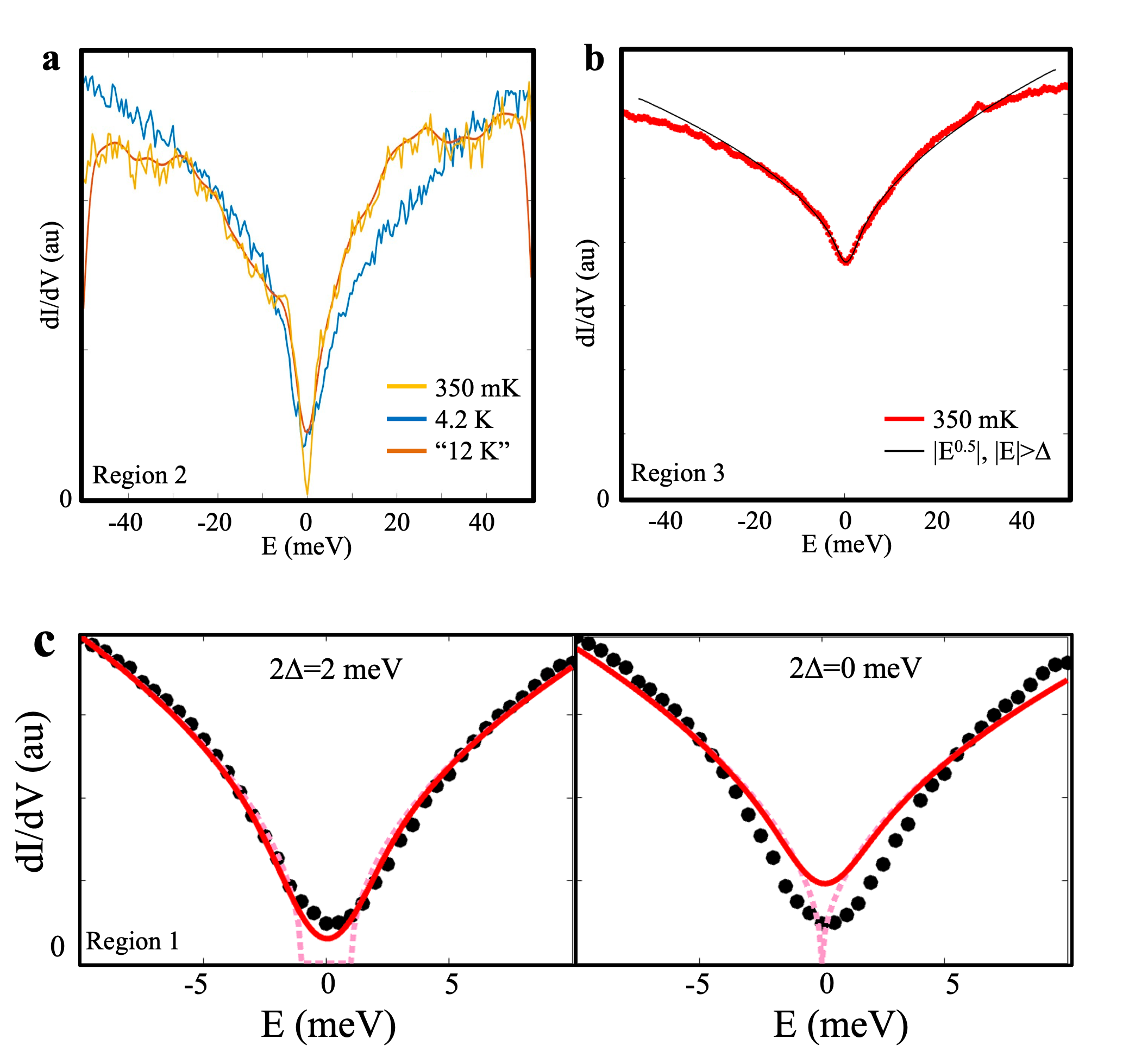}
\caption{\textbf{zero bias suppression in dI/dV spectrum. a} dI/dV spectrum measured at a location on the sample different than that shown in Fig.\ref{fig2} showing full zero bias suppression at base temperature of 350 mK that heals beyond thermal broadening at 4.2 K. \textbf{b} dI/dV spectrum measured at a third location showing a zero bias suppression with a finite residual density of states. \textbf{c} Fitting the vicinity of zero bias (same as in Fig.\ref{fig2}b) to a
square root energy dependence with slightly gapped versus nodal spectrum (dotted lines in left and right panels, respectively). The solid line further include smoothing by finite ac amplitude and finite temperature.} 
\label{figs-soft}
\end{figure}

In Fig.\ref{figs-soft}c we carefully examine the vicinity of zero bias of the dI/dV spectrum shown in Fig.\ref{fig2}b. The convex curvature of the dI/dV spectrum does not seem to extrapolate towards zero bias but rather to finite voltage intercept. We fit the low energy dI/dV spectrum to a phenomenological square root energy dependence both with and without a finite gap offest, $2\Delta$, on the left and right panels, respectively. The nominal square-root curves (black dotted lines) are further broadened by finite temperature of 400 mK 
and finite ac excitation of 1 mV (solid lines). This analysis suggests that a few meV gap is needed to accurately capture the measured zero bias dip. One origin for the square root energy dependence could thus be  the dispersion of three dimensional bands about that narrow gap in the DOS. Intriguingly, a 2 meV gap corresponds to a temperature scale of about 15 K, that agrees with the crossover temperature we have found in longitudinal resistivity and anomalous Hall conductivity.

\clearpage

\section{Saturated Anomalous Hall Conductivity}

\begin{figure}[ht]
\centering
\includegraphics[scale=0.5]{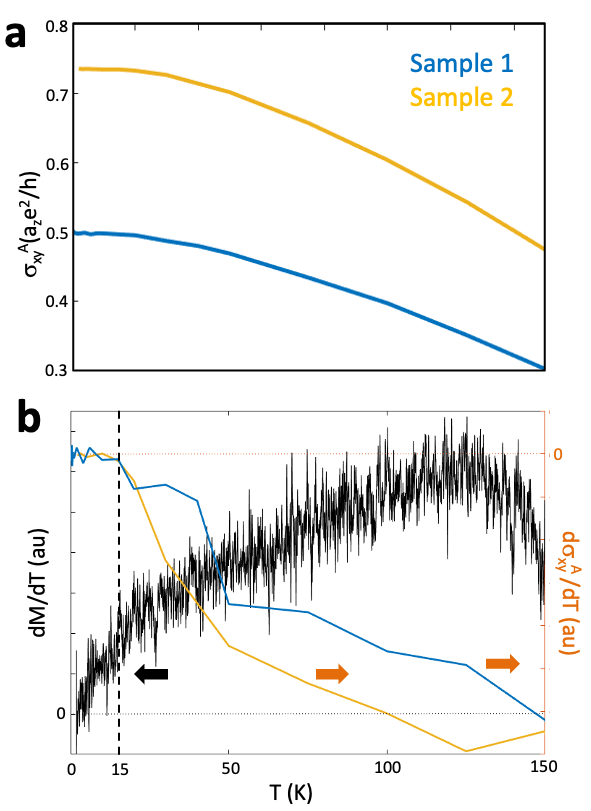}
\caption{\textbf{Saturation of the anomalous Hall conductivity versus lack of saturation of the magnetization. a} Anomalous Hall conductivity of two different Fe$_3$GeTe$_2$ samples presented versus linear temperature scale (Sample 1 is the same data presented in logarithmic temperature scale in Fig.\ref{fig1}f). \textbf{b} Derivative with respect to temperature of both the anomalous Hall conductivity of the two samples (blue and orange lines, right axis) and the magnetization of sample 1 (black line, left axis)}
\label{figs-sat}
\end{figure}

We carefully examine the low-temperature saturation we observe in the Anomalous Hall conductivity and compare it to the temperature dependence of the magnetization. First we show in Fig.\ref{figs-sat}a that the saturation is not unique to a specific sample by presenting two samples. Even though they show distinct absolute values values of $\sigma_{xy}^{A}$ at all temperatures, still they both seem to reach their maximal value and plateau below about 15 K. To better resolve the saturation we present in Fig.\ref{figs-sat}b their derivative with respect to temperature. In both sample the derivatives becomes zero below 15 K signifying lack of temperature dependence within that range. We compare this to the derivative of the measured magnetization with respect to temperature (black line) which in contrast remain finite at about 15 K and possibly approaches zero at the lowest temperatures measured, if at all. This signifies that the saturation in anomalous Hall conductivity does not result from the temperature dependence of the magnetization.

\clearpage

\section{Magnetization}
\begin{figure}[ht]
\centering
\includegraphics[width=0.98\linewidth]{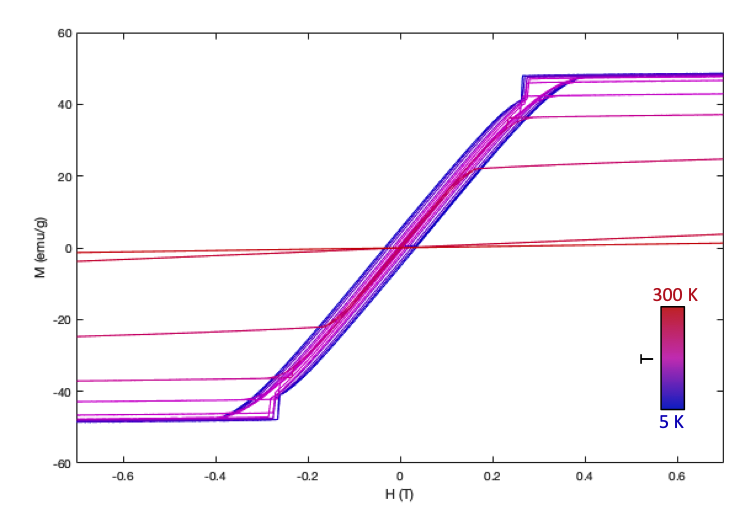}
\caption{\textbf{Magnetization.} Magnetization loops as a function of applied magnetic field at various temperatures between 5 and 300 K.}
\label{figs-mh}
\end{figure}

The ferromagnetic transition happens at 220 K as seen by the onset of a remnant magnetization and a hysteresis loop in Fig.\ref{figs-mh}. The profile of the hysteresis corresponds to the gradual polarization of ferromagnetic domains. A typical single Barkhausen jump occurs in the demagnetization branches for temperatures of 150 K and below.
The jumps increase in size with decreasing temperature and indicate the first domain magnetization reversal
of the fully saturated sample.

\clearpage

\section{Magneto-transport} \label{mag-trans}

\begin{figure}[ht]
\centering
\includegraphics[width=0.98\linewidth]{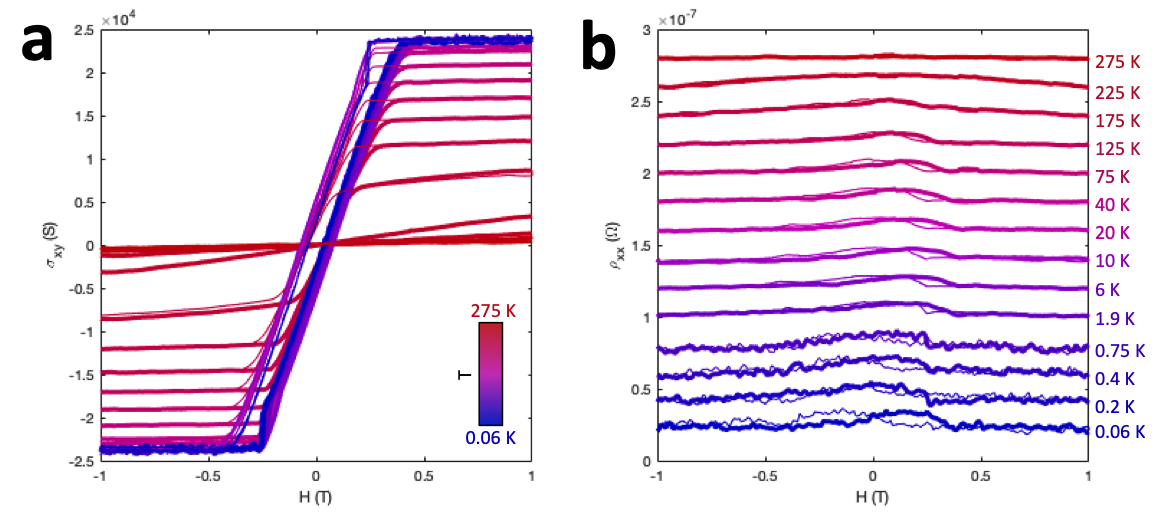}
\caption{\textbf{Magneto-transport. a} Magnetic field dependence of the Hall conductivity at various temperatures between 60 mK and 275 K showing a very large anomalous contribution and onset of hysteresis below the Curie temperature. \textbf{b} the longitudinal resistivity along the magnetic field cycles measured at corresponding temperatures shows a butterfly-like hysteretic pattern.}
\label{figs-rsh}
\end{figure}

The Hall conductivity dependence on the magnetic field, shown in Fig.\ref{figs-rsh}a, follows the trend of the magnetization (Fig.\ref{figs-mh}) including the hysteresis and the domain evolution. By linear extrapolation of the high field branches to zero field the anomalous Hall conductivity is extracted. From the slope of the classical Hall resistivity, $\rho_{xy}$, at high $B$ fields we also extract the carrier density in the sample, $n_e=-w/le B/\rho_{xy}$, where $w=300 \mu$m and $l=1$ mm are the sample width and length, respectively, and $e$ the electronic charge. From it we calculate the Fermi wavelength, $K_F=(3\pi^2n_e)^{1/3}$. The longitudinal resistivity, shown in Fig.\ref{figs-rsh}b is much less affected by the magnetization evolution in temperature. Still, a broad peak-like feature onsets where the hysteresis closes and the magnetization and Hall conductivity plateau. Those peaks may be associated with domain wall motion or with percolating edge conductance of a quantum Hall state.   

\clearpage

\section{Reproducibility}
\begin{figure}[ht]
\centering
\includegraphics[width=0.98\linewidth]{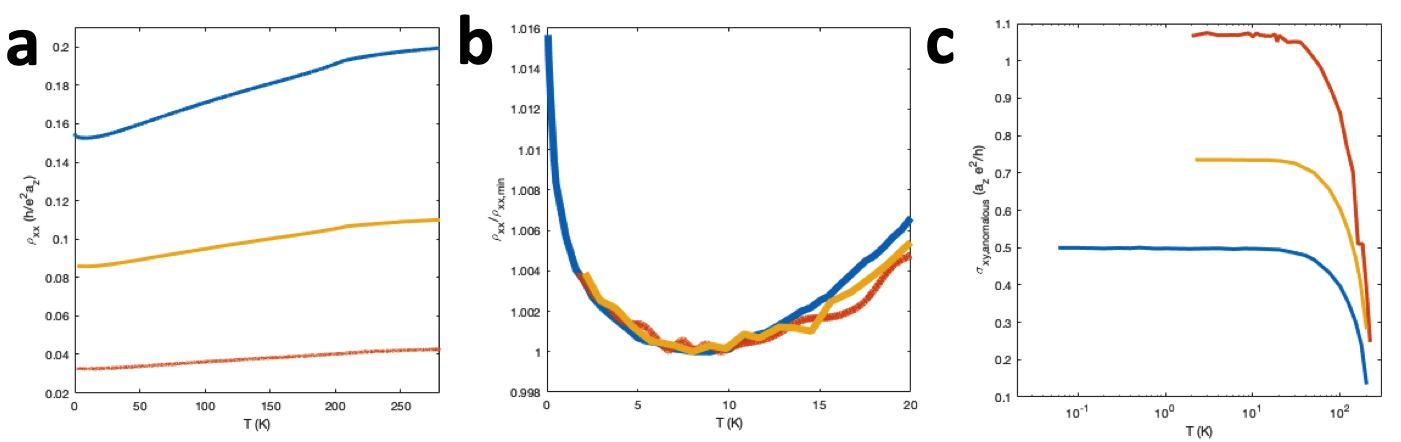}
\caption{\textbf{Reproducibility of the cross over temperature. a} Temperature dependence of the longitudinal resistivity for 3 different Fe$_3$GeTe$_2$ samples. \textbf{b} The low temperature part of the longitudinal resistivity after collapsing the room temperature values. \textbf{c} The temperature dependence of the anomalous Hall conductivity of the 3 samples.}
\label{figs-rst}
\end{figure}

We have examined three different samples in magneto-transport. The room temperature longitudinal resistivity of the most resistive sample is 5 times higher than that of the least resistive, as seen in Fig.\ref{figs-rst}a. This signifies different disorder content. Yet, after scaling the room temperature resistivity the three curves collapse and the cross-over temperature from metallic-like to resistive-like dependence occurs for all at about 15 K, as shown in Fig.\ref{figs-rst}b. Similarly, the  temperature below which the anomalous Hall conductivity saturates to its maximal value is the same across samples as seen in Fig.\ref{figs-rst}c, even though the saturation value itself varies between 0.5-1.05 quantum conductance per layer. 

\clearpage

\section{temperature evolution of the zero bias dI/dV dip}
\begin{figure}[ht]
\centering
\includegraphics[width=0.98\linewidth]{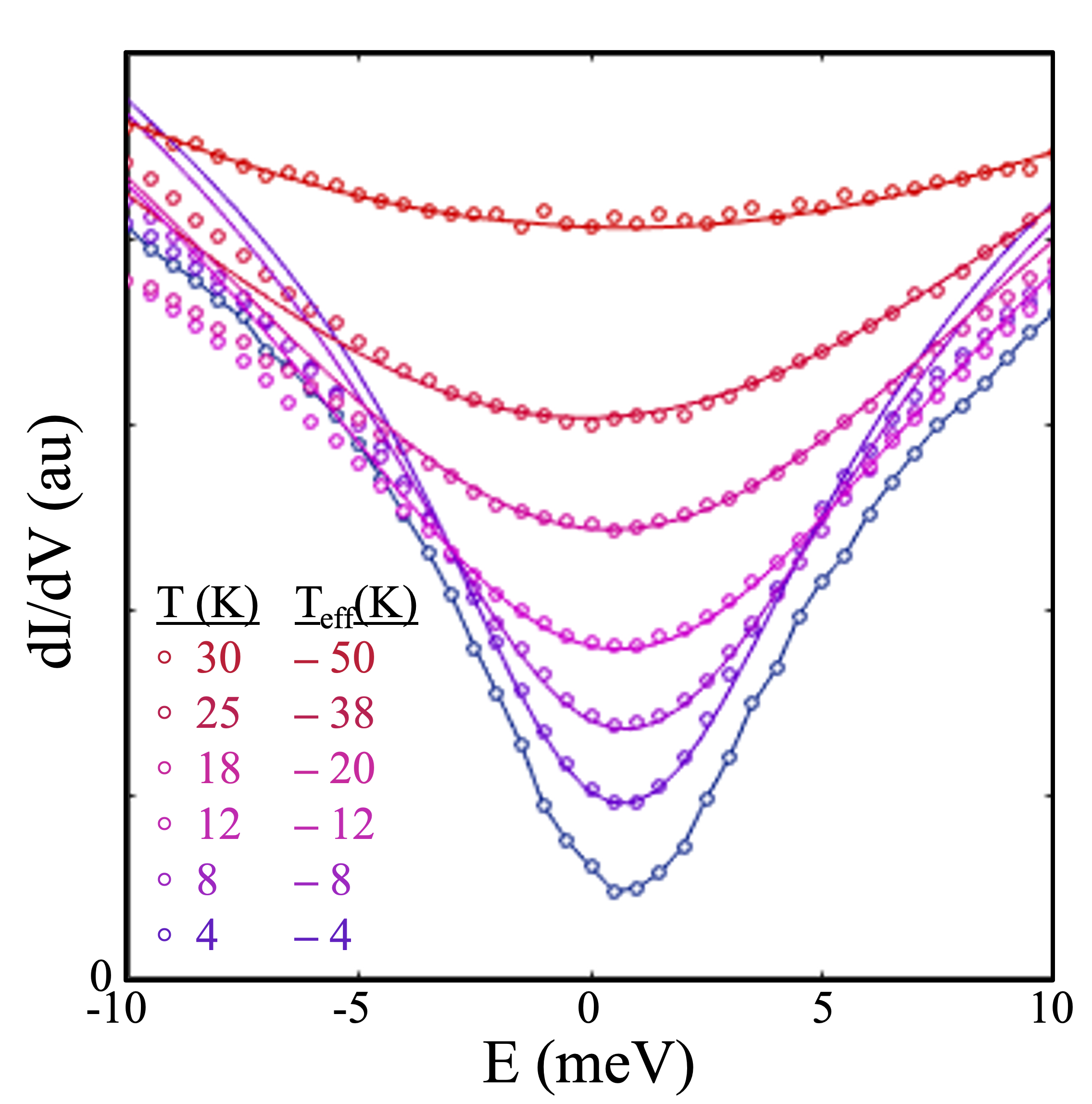}
\caption{\textbf{Temperature evolotion of the zero bias dip in dI/dV.} The data measured at varying temperature between 400 mK and 30 K (blue and red, respectively) and the result of thermally broadenning the lowest temperature data with a Fermi-Dirac distribution with an effective temperature.}
\label{figs-teff}
\end{figure}

When the temperature is varied from 400 mK to 30 K the zero bias dip we find in the dI/dV signal almost vanishes completely as seen in Fig.\ref{figs-teff}. We wish to distinguish between thermal broadening due to the Fermi-Dirac distribution from an evolution in temperature of the band structure. For that we take the lowest temperature data set measured at 400 mK and examine what effective temperature we need to plug into the Fermi-Dirac ditribution in order to capture the dI/dV curves measured at higher temperatures. We are able to obtain tight fits by this procedure as shown by solid lines. The extracted effective temperature is presented in Fig.\ref{fig3}.

\clearpage

\section{Multifractal computation method.}
The multifractality behaviour of a system is mainly understood by the signature plot known as singularity spectrum $f(\alpha(q))$ vs $\alpha(q)$. The singularity spectrum is generally an inverted parabola, with the points indicating the various fractal dimensions of the system. The calculation to obtain $f(\alpha(q))$ is carried out using box-counting method (see \cite{ChhabraJensen}). The maps were partitioned into boxes of various sizes $L= 2^n$. A series of normalized values of $\mu_i(q,L)$ was produced using positive and negative values of $q$, with incremental changes of $0.1$ across the range of variables.

\begin{equation}
\label{eqn:mf_mui}
\mu_i (q,L)=\frac{P_i^q(L)}{\sum_{i=1}^{N(L)}P_i^q(L)}
\end{equation}
In the above equation, $P_i(L)$ denotes the probability of density of states/differential conductivity present in each box of size $L$ and is  given by. 

\begin{equation}
\label{eqn:mf_pi}
P_i(L) = \frac{N_i(L)}{N_T}
\end{equation}

where $N_i(L)$ is the number of pixels that contains the density of states in the ith box, and $N_T$ is the total density of states in the spectroscopic map. The values in the singularity spectrum of multifractal $f(\alpha(q))$ were computed using

\begin{equation}\label{eqn:mf_faq}
f(\alpha(q)) = \lim_{L \to 0 } \frac{\sum_{i=1}^{N(L)}\mu_i (q,L) \log[(\mu_i (q,L))]}{\log L}
\end{equation}
and the $\alpha (q)$ values were obtained by,
\begin{equation}\label{eqn:mf_aq}
\alpha(q) = \lim_{L \to 0 } \frac{\sum_{i=1}^{N(L)}\mu_i (q,L) \log[P_i(L)]}{\log L}
\end{equation}

By varying values of q in steps of $0.1$, the values of $f(\alpha(q))$ and $\alpha (q)$ were extracted from the slopes of the numerators of equations \ref{eqn:mf_faq}, \ref{eqn:mf_aq} and their denominator, $\log L$. The values of $L$ varied from $2^0$ to the maximum ($2^n$) pixels available in the spectroscopic map.

The $\alpha_0$ and $\Delta\alpha$ values are extracted from the plot of $f(\alpha(q))$ vs $\alpha(q)$. The $\alpha_0$ is taken at $q=0$ and $\Delta\alpha$ is obtained from the difference in the points at $q=-1$ to $q=+1$.

\subsection{Estimation of the expected fractal dimension} \label{supp-frac-r}

For non interacting quasi 2D systems for which $\sigma{xy}\ll\sigma{xx}$, the expected deviation from the physical dimension can be estimated from the following relation, that relates the fractal dimension to the longitudinal conductance \cite{efetov_1996} :

\begin{equation}
d_2-2=\frac{2}{2 \pi^2 \beta_0 \hslash \nu D}
\label{eqn:d2}
\end{equation}

Here $d_2$ is the fractal dimension and $\beta_0 =2$. Substituting $D\nu_0=\frac{\sigma_{xx}}{e^2}$ we obtain 

\begin{equation}
\label{eqn:d22}
2-d_2=\frac{2 e^2}{\pi\beta_0 h\sigma_{xx}}, \quad d_2=2-\frac{\rho_{xx} e^2}{\pi h}
\end{equation}

For $\rho_{xx}= 330 \frac{\mu\Omega}{Cm}\approx2200 \frac{\Omega}{layer}\approx0.085\frac{h}{e^2}$ obtained in our measurements we get $d_2=1.973$. Converting to $\alpha_0$ using  $d_2=2-\gamma$, $\alpha_0=2+\gamma$, we obtain $\alpha_0 \approx$2.02 in consistent with what we extract from the dI/dV maps.

\section{Kondo transition}
\begin{figure}[ht]
\centering
\includegraphics[width=0.98\linewidth]{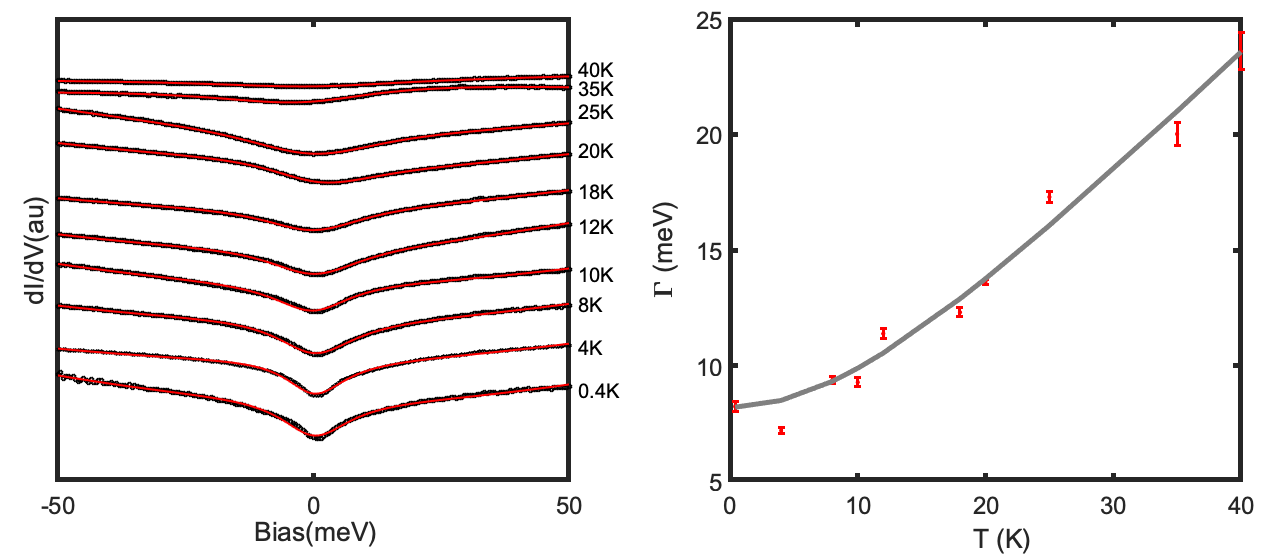}
\caption{\textbf{Kondo analysis. (a)} dI/dV spectra measured at various temperatures (black dots, shifted for clarity), each fitted with a Fano line shape (red lines) \textbf{(b)} $\Gamma$ values *red symbols) extracted from the fits in (a) fitted with a formula for a Kondo transition (solid line) yielding $T_K \approx$ 31 K.}
\label{figs-kondo}
\end{figure}

We examine the Kondo state origin for the zero bias dip in the tunneling density of states. While tunneling into a Kondo impurity should reveal a low bias resonance, interference with a tunneling path to the surrounding metal that screens it would result in a Fano line shape: $dI/dV(E)=(E+q\Gamma)^2/(\Gamma^2+E^2)$. Fitting the measured dI/dV spectra taken at increasing temperatures above a linearly dispersing background with a Fano spectra (Fig.\ref{figs-kondo}a, black versus red lines, respectively) shows good agreement. Fits at all temperatures yield vanishingly small tunneling ratio, $q\approx0.1$, suggesting that we tunnel predominantly into the screening metal rather than to the Kondo impurities. This is reasonable considering that we tunnel to the Te termination while the Kondo impurities are thought to originate from the buried Fe layers. 

The individual Kondo fits allow us to extract the temperature dependence of the resonance width, $\Gamma(T)$, plotted in Fig.\ref{figs-kondo}b. There we find that the high temperature linear dependence of the width expected due to thermal broadening saturates below about 10 K signifying a Kondo transition. Fitting these with a statndard Kondo formula for the crossover:
\begin{equation}
   \Gamma(T) \propto [(\pi K_B T)^2+2(K_BT_K)^2]^{1/2}
\end{equation}
yields a Kondo transition of $T_K \approx 31$ K in qualitative agreement with the low temperature crossover found in longitudinal resistivity $\rho_{xx}$.

We compare the phenomenology of the Kondo and Altshuler-Aronov efects relevant to our findings in the table below:

\begin{table}
\begin{center}
\caption{Comparison of phenomenology of Kondo and Altshuler-Aronov effects} \label{tbl-kondo-aa}

\begin{tabularx}{0.98\textwidth} { 
  | >{\raggedright\arraybackslash}X 
  | >{\raggedright\arraybackslash}X 
  | >{\raggedright\arraybackslash}X | }

 \hline
 phenomenon & Kondo & Altshuler-Aronov \\ 
 \hline
 Logarithmic increase in longitudinal resistivity with decreasing temperature & 
 \checkmark Commonly found in metals with magnetic impurities \cite{kondo1964resistance, de1934electrical}. In the dense impurity case or in Kondo lattices a drop in resistivity is expected below the coherence temperature that we do not observe \cite{yang2008scaling,lavagna1982resistivity}. & 
 \checkmark Predicted to give rise to logarithmic correction and observed is several systems (like graphene) \cite{PhysRevLett.44.1288,kozikov2010electron}. \\ 
 \hline
 Zero bias suppression in the dI/dV spectrum & 
\checkmark The Fano line shape in the tunneling dI/dV appears as a dip in the limit of dominant tunneling to the metal and negligible tunneling to the magnetic impurities \cite{schiller2000theory, aguiar2007kondo}. & 
\checkmark Expected due to the low conductivity and their poor screening in disordered metals \cite{AltshulerAronov1979,levitov1997semiclassical}. \\ 
 \hline
Multifractality &
$?$ No known prediction. Not well studied. &
\checkmark Predicted in systems with criticality (for instance chiral 2D or at higher dimensions) \cite{janssen1994multifractal, Evers2008} \\
 \hline
Anomalous Hall conductivity &
$?$ Some discussion \cite{RAMAKRISHNAN1985493} and observations in Kondo lattice systems \cite{xu2024tunable} &
$?$ No consistent picture on the mutual dependence \cite{mitra2010localization,mitra2007weak} \\
 \hline

\end{tabularx}
\end{center}
\end{table}
\label{sm-kondo}

\clearpage

\section{Altshuler-Aronov effect in layered materials} \label{sm-aa}

We present details of the calculations underlying Eq.\ (1) of the main text. The correction to the tunneling density of states at energy $\epsilon$ and temperature $T$ due to disorder and interactions is given by \cite{PhysRevLett.44.1288,AltshulerAronov1979}, 
\begin{equation}
    \delta\nu(\epsilon,T) = - \nu \, \mathrm{Im}\int_0^\infty \frac{d\omega}{2\pi} \int \frac{d\mathbf{q}}{(2\pi)^3} U(\mathbf{q},\omega)  \frac{\tanh\frac{\omega+\epsilon}{2T}+\tanh\frac{\omega-\epsilon}{2T}}{(-i\omega + Dq_\parallel^2)^2}
\end{equation}
This expression assumes electronic diffusion and treats electron-electron interactions perturbatively. Here, we assume that diffusion is only occurring within the 2D layers (diffusion constant $D=k_F l/2$), while the screened Coulomb interaction 
\begin{equation}
U(\mathbf{q},\omega) = \frac{4\pi e^2}{q^2 + \kappa^2 \frac{Dq_\parallel^2}{|\omega| + D q_\parallel^2}}    
\end{equation}
is three dimensional, both in the field lines going into the third dimension and in the screening electrons occupying three dimensions. The noninteracting density of states is $\nu = \nu_{2d}/d$ with $\nu_\mathrm{2d}=m/2\pi$. The Thomas-Fermi wavevector $\kappa$ is $\kappa = 4\pi e^2\nu$. The wavevector $\mathbf{q}=(\mathbf{q}_\parallel,q_z)$ has components parallel and perpendicular to the layers. We set $\hslash=k_B=1$ here. 

At zero temperature, the tunneling density of states simplifies to 
\begin{equation}
    \delta\nu(\epsilon) = - 2\nu \mathrm{Im}\int_{|\epsilon|}^\infty \frac{d\omega}{2\pi} \int \frac{d\mathbf{q}}{(2\pi)^3} U(\mathbf{q},\omega) \frac{1}{(-i\omega + Dq_\parallel^2)^2}.
\end{equation}
The integration over $q_z$ is only over the screened Coulomb interaction. Performing the integration, one obtains 
\begin{equation}
    \delta\nu(\epsilon) = - 2\nu \mathrm{Im}\int_{|\epsilon|}^\infty \frac{d\omega}{2\pi} \int \frac{d\mathbf{q}_\parallel}{(2\pi)^2} \frac{2\pi e^2}{\sqrt{q_\parallel^2 + \kappa^2 \frac{Dq_\parallel^2}{-i\omega + Dq_\parallel^2}}} \frac{1}{(-i\omega + Dq_\parallel^2)^2}.
\end{equation}
Changing variables to $y=Dq_\parallel^2$ yields
\begin{equation}
    \delta\nu(\epsilon) =  \frac{\kappa^2}{4\pi\sqrt{D}} \mathrm{Im}\int_{|\epsilon|}^\infty \frac{d\omega}{2\pi} \int_0^\infty \frac{dy}{\sqrt{y}} \frac{1}{\sqrt{\omega  + i(y+D\kappa^2) }(\omega + i y)^{3/2}}.
\end{equation}
The integration over $\omega$ can be performed exactly and yields
\begin{equation}
    \delta\nu(\epsilon) =  \frac{1}{2\pi D^{3/2} } \mathrm{Re}\int_0^\infty \frac{dy}{\sqrt{y}}\left\{1 - \sqrt{\frac{|\epsilon| + i (y + D\kappa^2)}{|\epsilon| + iy} } \right\}.
\end{equation}
The integral converges in the regions $y\ll |\epsilon|$ and $y\gg D\kappa^2$, so that it is dominated by the logarithmic divergence in the region $|\epsilon| \ll y \ll D\kappa^2$. Retaining only this large logarithmic contribution, we obtain
\begin{equation}
    \delta\nu \simeq - \frac{\kappa}{2\pi D} \ln \frac{D\kappa^2}{|\epsilon|},
\end{equation}
which underlies Eq.\ (1) of the main text. 

\printbibliography[title=Supplementary References]

\end{refsection}


\end{document}